\documentclass[runningheads]{llncs}
\usepackage{eccv}

\usepackage{eccvabbrv}

\usepackage{graphicx}
\usepackage{booktabs}
\usepackage{makecell}
\usepackage{ifthen}

\usepackage[accsupp]{axessibility}  % Improves PDF readability for those with disabilities.

\usepackage[pagebackref,breaklinks,colorlinks,citecolor=eccvblue]{hyperref}
\usepackage{orcidlink}
\usepackage{algorithm}
\usepackage{algpseudocode}
\usepackage{bm}
\usepackage{amsmath}
\usepackage[symbol]{footmisc}

\newcommand{\secref}[1]{Section~\ref{sec:#1}}

\newcommand{\figref}[1]{Fig.~\ref{fig:#1}}
\renewcommand{\eqref}[1]{Eq.~(\ref{eq:#1})}
\definecolor{MyDarkRed}{rgb}{0.8,0.02,0.02}

\newcommand{\vv}{\mathbf{v}}

\newcommand{\VV}{V}

\newcommand{\ee}{\mathbf{e}}
\newcommand{\ec}{\mathbf{e}^\text{cloth}}
\newcommand{\eb}{\mathbf{e}^\text{body}}
\newcommand{\esc}{\mathbf{e}^\text{self-col}}

\newcommand{\xx}{\mathbf{x}}
\newcommand{\FF}{F}

\begin{document}

\title{SENC: Handling Self-collision in Neural Cloth Simulation} 

\titlerunning{SENC}

\newcommand{\equalcontrib}{\textsuperscript{*}}
\newcommand{\correspond}{\textsuperscript{$\dagger$}}

\footnotetext[1]{Equal contribution.}
\footnotetext[2]{Corresponding author.}
% \renewcommand{\thefootnote}{\fnsymbol{footnote}}
% \footnotetext[1]{Equal contribution.}
\author{Zhouyingcheng Liao\equalcontrib\orcidlink{0009-0002-6525-1372} \and Sinan Wang\equalcontrib\orcidlink{0009-0005-9322-2351} \and Taku Komura\correspond\orcidlink{0000-0002-2729-5860}}

\authorrunning{Z. Liao et al.}

\institute{The University of Hong Kong}
% \url{http://www.springer.com/gp/computer-science/lncs}}
\maketitle

\centerline{\url{https://zycliao.github.io/senc}}

% $\xx$: vertex feature \\
% $\ee$: edge feature \\
% $\vv$: vertex position (single) \\
% $\VV$: all vertices \\
% $\FF$: all faces \\
% $\vc_i$ or $\vv_i$: cloth vertex i \\
% $\VB_i$: all body vertices \\
% $\eb$: body-cloth edge \\
% $\esc$ : self-collision edge

\begin{figure}
    \centering
    \includegraphics[width=0.85\textwidth]{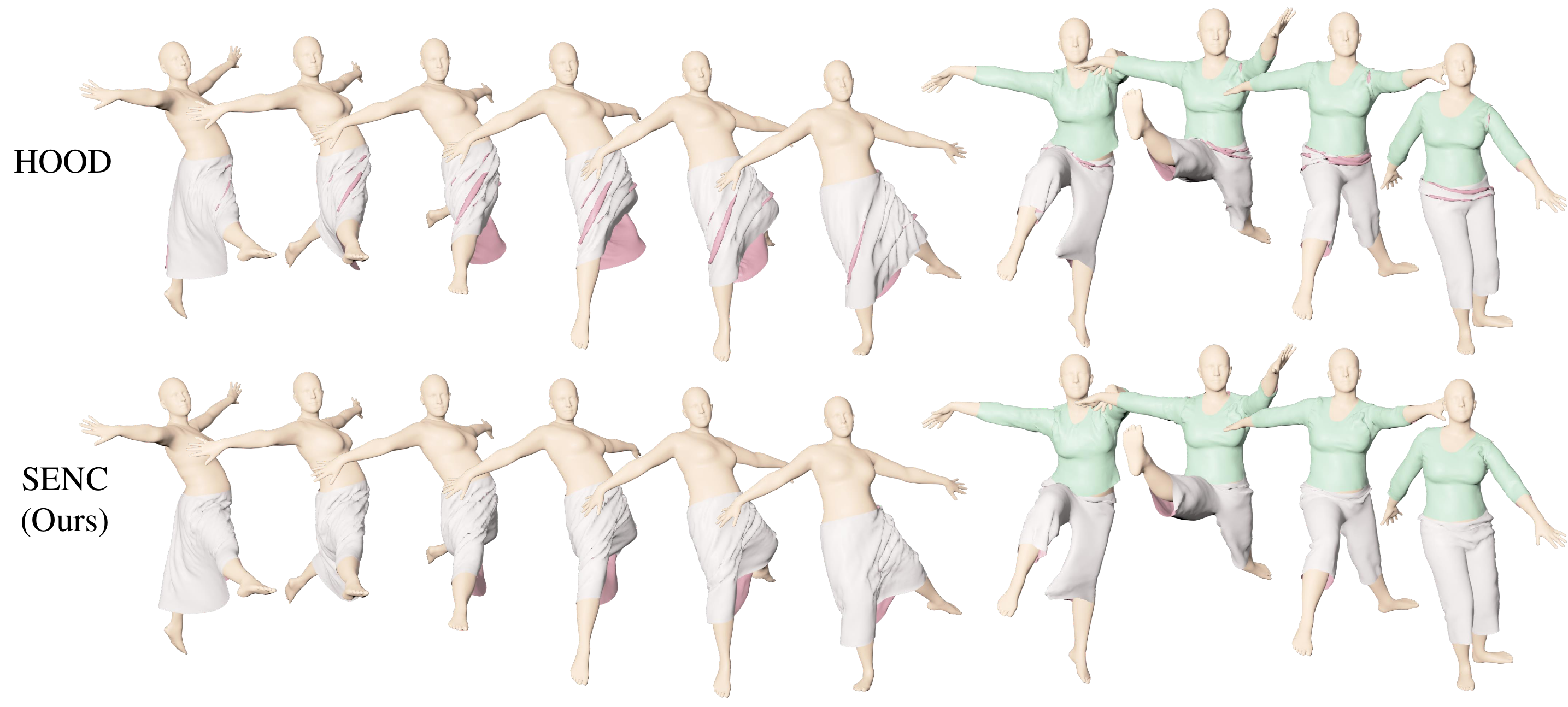}
    \caption{Our method effectively addresses cloth self-collision, compared to existing state-of-the-art neural cloth simulator~\cite{grigorev2023hood}. Note the inner side of the cloth is painted pink.}
    \label{fig:teaser}
\end{figure}
\vspace{-0.7cm}

\begin{abstract}
We present SENC, a novel self-supervised neural cloth simulator that addresses the challenge of cloth self-collision. This problem has remained unresolved due to the gap in simulation setup between recent collision detection and response approaches and self-supervised neural simulators. The former requires collision-free initial setups, while the latter necessitates random cloth instantiation during training.
To tackle this issue, we propose a novel loss based on Global Intersection Analysis (GIA). This loss extracts the volume surrounded by the cloth region that forms the penetration. By constructing an energy based on this volume, our self-supervised neural simulator can effectively address cloth self-collisions. Moreover, we develop a self-collision-aware graph neural network capable of learning to handle self-collisions, even for parts that are topologically distant from one another.
Additionally, we introduce an effective external force scheme that enables the simulation to learn the cloth's behavior in response to random external forces. 
We validate the efficacy of SENC through extensive quantitative and qualitative experiments, demonstrating that it effectively reduces cloth self-collision while maintaining high-quality animation results.

\end{abstract}

\section{Introduction}

Self-supervised neural cloth simulation is attractive in the sense that the system can train itself and does not require the user to prepare ground truth garment data, which is usually very expensive to acquire.
% \sinan{Also, unlike supervised methods, self-supervised approaches are more physically correct because they optimize the physical losses like traditional numerical simulation.}
Among self-supervised techniques, methods that are trained for specific garments~\cite{santesteban2022snug, bertiche2022neural} or those that can generalize to arbitrary garments exist~\cite{grigorev2023hood}. 
\par
Despite the advancements in such self-supervised neural cloth simulation technologies, a critical impediment to the generation of realistic and precise animations remains unaddressed: \emph{cloth self-collision}, which is a very common type of animation artifact that can happen in various conditions, such as the self-penetrations that occur in clothing due to close contact areas, like under the arms when the arms are pressed against the body and overlayed skirts (shown in Figure~\ref{fig:teaser} and Figure~\ref{fig:comparison}).
\par
% 2. Difficulties of neural self-collision
%    why traditional methods can't apply here
Although a variety of collision detection and response techniques~\cite{baraffW98,shi2023unified,teran2005robust,li2020incremental} have been proposed for traditional physical-based simulation, they cannot be easily applied to self-supervised neural cloth simulators due to the special treatment needed.
% IPC cannot apply here
%For example, the incremental potential contact~\cite{li2020incremental}, the state-of-the-art collision detection and response technique, cannot produce gradients to resolve the collisions when self-collision already exists.
%However, most neural cloth simulators predict the cloth state in one forward pass, which cannot avoid the occurrence of self-collision.
%Also, its collision energy increases to infinity when the colliding pairs approach each other, leading to gradient explosion during training.
For example, the incremental potential contact (IPC)~\cite{li2020incremental}, the state-of-the-art collision detection and response technique, uses a barrier method to prevent collisions. It assumes the simulated object starts from a non-colliding state, and the collision energy increases to infinity 
when the colliding pairs approach each other, which may lead to gradient explosion if used in neural simulation. 
Such a characteristic makes it unsuitable for neural cloth simulators, which randomly instantiate the state of the garment during the training process.  
Other techniques that form a penalty function based on edge-edge collisions of two consecutive frames~\cite{baraffW98,shi2023unified} are also not suitable for the same reason,  
as they also require starting from a collision-free state. 
%Finally, methods based on signed distance functions (SDF) are also not suitable as the forces produced by the gradient of the SDF only locally push the particles to the nearest surface, 
%and  do not coordinate to resolve the penetration in a global manner;  leading to local minima; {e.g., when one part is deeply penetrated into another, the middle part may be stuck inside the mesh when two ends of the penetration volume is dragged towards two reverse directions.}
Finally, methods based on signed distance functions (SDF)~\cite{teran2005robust} are not suitable for our purposes, as the forces generated by the gradient of the SDF only locally push the particles towards the nearest surface, resulting in local minima. For instance, in cases where one part deeply penetrates another, the middle section may become trapped inside the mesh when the two ends of the penetration volume are pulled in opposite directions.

%\sinan{especially when the penetration is very large and approaches another part of the mesh, then that end of the penetration will be dragged to that part of the mesh, instead of the original part of the mesh that it penetrates in.}

%Thus, such methods cannot be used for training neural cloth simulators.
% For example, the collision energy that is applied in incremental potential contact \cite{li2020incremental} increases to infinity when the colliding pairs approach each other to remove any risk of penetrations. Although it has been shown that such an energy can reproduce realistic deformation of deformable objects that goes under severe deformation and intense collisions, it cannot produce gradients to resolve the collisions when self-penetration already exists.  Such a scheme requires simulating the interaction of garments from states where no penetrations exist, which is not the case for self-supervised networks where the batch starts from random conditions where self-collision can exist. 
% post-processing cannot apply here
%Another possible branch of solutions is to post-process the self-collision after every network forward pass using methods like untangling cloth~\cite{baraff2003untangling}. However, processing self-collision after predicting the cloth position corrupts the energy minimization achieved before, altering its physical accuracy. Moreover, such processing introduces a high computational cost, negating the speed advantage of neural cloth simulation.
\par
% 3. difficulty 2, it's hard to define the penetration side
%The other difficulty lies in the fact that the self-collision of cloth-like objects is different from the collision between the cloth and other closed objects, or the collision between different closed objects, where the penetration side is clearly defined and the collision energy can be easily designed. Instead, it is topologically ambiguous to define the penetration side for cloth-like objects due to their thin nature.
\par
% 4. we propose SENC
To overcome these difficulties, we propose SENC, \ie, handling \textbf{SE}lf-collision in \textbf{N}eural \textbf{C}loth simulation, which composes a novel self-collision loss
and a self-collision-aware graph neural network.
%Once trained, it predicts the cloth state autoregressively for garments of arbitrary topologies and mesh connectivity, with minimal self-collision.
%Our model is trained in a self-supervised way by several physical energy-based terms, of which the novel self-collision loss is the key to preventing self-collision. 
The self-collision loss is based on the 
volume surrounded by the self-penetrating area of the cloth,
which is computed by Global Intersection Analysis (GIA)\cite{baraff2003untangling}.
The idea of GIA is to analyze the boundaries of self-collisions, i.e., intersection paths, and identify vertices within the intersection paths.
The gradient of the volume naturally forms a force that drives the garment vertices in the direction that resolves the collision.
% Thanks to the global nature of the analysis, our loss is less likely to lead to local minima, %\sinan{where the penetrating part may be stuck inside the mesh because two ends of the penetration volume is dragged towards two reverse directions}, 
% which can happen with SDF-based loss.
%After GIA, the untangling cloth \cite{baraff2003untangling} applies forces to the cloth vertices within the penetration regions in the direction that self-collisions are resolved, i.e., pulls them together.
%This process occurs after the time integration step and is not suitable for our objective. We aim for the predicted garment to be free of self-collisions without the need for any additional procedures. 
%Thus we propose a novel self-collision loss based on the volume that is surrounded by the self-penetrating area of the cloth.  

%Moreover, our self-collision loss could be further applied to solve the self-collision of other deformable objects, such as 3D characters.

Our method is developed upon the Graph Neural Network~\cite{scarselli2008graph} (GNN) architecture, which has shown excellent performance in learning cloth dynamics~\cite{pfaff2020meshgraphnet,sanchez2020learning,grigorev2023hood}, but has failed to consider self-collisions.
Previous methods~\cite{pfaff2020meshgraphnet,sanchez2020learning,grigorev2023hood} form graphs according to the topological connectivity of the mesh, where the nodes correspond to the vertices of the mesh, and the edges represent their connections.
Although such graphs are effective in predicting the dynamics of clothing based on the local deformation of the fabric caused by the interaction of topologically adjacent areas, 
%adjacent areas and propagates based on topological connections of the mesh,
these structures cannot be used to prevent cloth self-collision, as many self-collisions happen between areas with a large topological distance.
To address this issue, we propose the self-collision-aware GNN, where we construct additional edges based on the spatial distance of vertices, which effectively prevents cloth self-collision.
In addition, our model can model variable external forces, allowing the users to provide external forces e.g. based on wind, to increase the dynamic behavior of the garment.
\par
% experiments
 We examine our scheme in various types of 
 garments, including t-shirts, pants, long sleeves, skirts, and dresses. 
 The experiments show that our approach can significantly reduce the amount of self-collisions compared to existing state-of-the-art methods (see \figref{teaser}). 
% scenarios that are mentioned above: \sinan{CHECK}including a garment worn by a character that 
% conducts various dynamic movement, garments with multiple layers worn by such characters, and a piece of cloth 
% dynamically folded in multiple layers. Our scheme can robustly avoid penetrations between the garments to produce 
% plausible deformation. \sinan{CHECK}Our scheme is applicable for frameworks that require specific garment connectivities or those 
% that generalize to arbitrary connectivities.
\par
% 7. summary
%In summary, we propose SENC, a neural cloth simulator that predicts realistic clothing dynamics for garments of arbitrary mesh connectivities with minimal self-collision, thanks to the self-collision-aware graph neural network and a novel self-collision loss.
In summary, the contribution of our paper can be summarized as follows: 
\begin{itemize}
\item A novel self-collision loss based on GIA that reduces self-collisions on self-supervised neural cloth simulation and
\item a self-collision-aware graph neural network that also allows users to apply external forces to the garment.
\end{itemize}

\section{Related Work}
In this section, we begin by reviewing cloth simulation, a foundational technique that has laid the groundwork for animating realistic garment deformations. We then explore neural cloth simulation, an innovative approach that has emerged as machine learning techniques have matured. Finally, we delve into collision detection and response techniques, which constitute this research's primary focus.

\subsection{Cloth Simulation}
Clothes are mostly simulated as a Lagrangian model where mass particles are connected to each other through elastic rods. The large time step is first achieved by using implicit time integration, as proposed in \cite{baraffW98}, which can reproduce plausible deformation of the clothes with realistic wrinkles and folds. 
Additionally, various methods based on mass springs \cite{Liu:2013:FSM}, finite element models \cite{sperl2020hylc}, projective dynamics \cite{10.1145/3527660}, yawn-level models \cite{sperl2020hylc, kaldor2008simulating}, and material point methods \cite{jiang2017anisotropic} have been proposed to realistically simulate the dynamics of the cloth. Despite their precision and the realism they offer, numerical simulation schemes can incur significant computational costs.
% A force model is proposed in \cite{baraffW98} that can reproduce plausible deformation of the clothes with realistic wrinkles and folds with a large time step. Despite their precision and the realism they offer, numerical simulation schemes can incur significant computational costs.

% Implicit integration is an essential component for animating cloth deformation in large steps.  Previous explicit integration results in small time steps
% that require many iterations for simulating the cloth deformation even only for a short amount of time. 
 
% Modern finite element approaches are based on incremental pontential contact that computes the comute the deformation of the cloth by evaluating the 
% Hessian and gradient of the energy, solving a linear system for computing the direction the particles evolve, and finally finding a step size that the 
% energy is minimized.   This approach has a good match with physics-informed neural networks that can be self-supervised without preparing a large amount of
% training data as the defined energy can be directly applied as the loss function of the neural network that predicts the next state of the cloth given
% the current state and external forces.

\subsection{Learning Garment Dynamics}
For offline film scenes, expensive numerical simulations are feasible, but there are cases like real-time games that demand faster solutions. Neural cloth simulation meets this need effectively. 
Some researchers try to learn cloth dynamics from the pose and/or shape of the 3D human and predict the clothing deformation in the unposed space~\cite{guan2012drape, wang2019learning,ma2020learning, patel2020tailornet, santesteban2019learning}. To finally deform the garment, these methods rely on linear blend skinning. Thus, they only suit those garments that closely stick to the body. The assumption that the cloth deformation is conditioned on the human pose also makes the result of these papers lack dynamics.
DeePSD~\cite{bertiche2021deepsd} trains a neural network to predict the cloth skinning weight from the canonical shape.
SSCH~\cite{santesteban2021self} introduces a diffused human model to project the ground truth cloth data into the canonical space for better learning in the canonical space.
These methods enable the deformation of looser garments.
However, all these methods require supervised learning with a great amount of ground truth data, and they often fall short of accurately capturing the true physical constraints of garments. 
In response, self-supervised learning strategies incorporating physical loss functions, such as SNUG \cite{santesteban2022snug}, ReFU~\cite{tan2022repulsive}, PBNS \cite{bertiche2020pbns}, 
 NCS~\cite{bertiche2022neural}, and GenSim~\cite{tiwari2023gensim} have emerged. These techniques strive to more faithfully mirror the underlying physical principles. Notably, the introduction of graph neural networks by HOOD \cite{grigorev2023hood} has demonstrated a robust capability for simulating realistic garment deformations. Methods combining supervised and unsupervised losses like GarSim \cite{tiwari2023garsim} also appear. 
 Recent diffusion-based methods~\cite{wang2024disentangled,yu2024surf} generate diverse types of garments, but they lack physical plausibility.
 CaPhy~\cite{su2023caphy} and ULNeF \cite{santesteban2022ulnef} try to resolve collisions between different layers of garments.
Yet, a critical issue that persists across all these approaches is their inability to address garment self-collision, a complex challenge that remains unresolved due to various previously outlined factors.
ClothCombo~\cite{lee2023clothcombo}, a quasistatic multilayer system but not a dynamic animation system, applies a simple repulsion loss to separate close vertices. Such forces cannot resolve self-collisions after it has already occurred. Concurrently, ContourCraft \cite{grigorev2024contourcraft} uses the contour length of the self-intersection as the unsupervised loss, which may even increase self-collisions when the true resolving direction is opposite to the direction of reducing the contour length of self-collision.
% complex self-intersection as can easily get stuck in local minima.
Our method effectively addresses this issue, offering seamless integration into existing neural cloth simulation frameworks, thus advancing the state-of-the-art in realistic neural cloth simulation.

\subsection{Collision Detection and Response}
Specialized acceleration data structures, such as Bounding Volume Hierarchies (BVH) \cite{ericson2004real} for spatial division, form the foundation of collision detection techniques. Various algorithms for constructing BVH, including Top-Down \cite{wald2007fast, hendrich2017parallel, lauterbach2009fast}, Bottom-Up \cite{walter2008fast, gu2013efficient}, and Incremental \cite{goldsmith1987automatic, bittner2015incremental} construction, together with Linear BVH \cite{pantaleoni2010hlbvh, karras2012maximizing}, have been proposed. Additionally, hardware support can be designed to further enhance the speed, as proposed in \cite{doyle2012hardware, doyle2013hardware, doyle2017evaluation}. For more details about BVH and its related work, we refer readers to \cite{meister2021survey}.
For modern cloth simulation methods using implicit time integration, Continuous Collision Detection (CCD) is used. It computes the maximum time step during the line search to prevent collision. Examples include CCD \cite{wang2021large} and additive CCD (ACCD) \cite{li2020codimensional}.
\par
After the collision has been detected, an appropriate collision response is required to handle and resolve the collision. For those methods using implicit time integration, the collision can be designed as some penalty energy and integrated into the optimization framework. Its recent representatives are Incremental Potential Contact (IPC) \cite{li2020incremental} for deformable objects and Codimensional IPC (C-IPC) \cite{li2020codimensional} for cloth-like thin structures. However, as previously mentioned, these approaches primarily aim to prevent collisions before they occur, which does not meet our requirements. Other methods capable of addressing collisions as they occur, such as Untangling Cloth \cite{baraff2003untangling}, or those involving local adjustments \cite{provot1997collision, bridson2002robust, harmon2008robust}, and with volume-preserving impulses \cite{sifakis2008globally}, are also not suitable due to their need for additional post-processing, therefore cannot be integrated into the framework of neural cloth simulation.
% \par
% In neural cloth simulation, there are also existing relevant methods like ULNeF \cite{santesteban2022ulnef}, ReFU \cite{tan2022repulsive},  CaPhy \cite{su2023caphy}, and ContourCraft \cite{grigorev2024contourcraft}. However, ULNeF \cite{santesteban2022ulnef} cannot handle complex poses and is only validated by T-pose. ReFU \cite{tan2022repulsive} is limited by the accuracy of the approximated human body signed distance function. CaPhy \cite{su2023caphy} relies on human body templates and therefore cannot easily deal with loose garments like long skirts. Concurrently, ContourCraft \cite{grigorev2024contourcraft} uses the contour length of the self-intersection as the unsupervised loss, which may even increase self-collisions when the true resolving direction is opposite to the direction of reducing the contour length of self-collision.

% In modern cloth simulation, with the implicit time integration mentioned above, collision handling is often treated as an optimization problem, where people compute the maximum time step during the line search to prevent the collision before its occurrence. Examples include Continuous Collision Detection (CCD) \cite{wang2021large}, Incremental Potential Contact (IPC) \cite{li}

% Traditional collision handling techniques involve two stages: collision detection and response. The collision is first detected and then resolved by appropriate response.

\begin{figure}[tb]
  \centering
  \includegraphics[width=\textwidth]{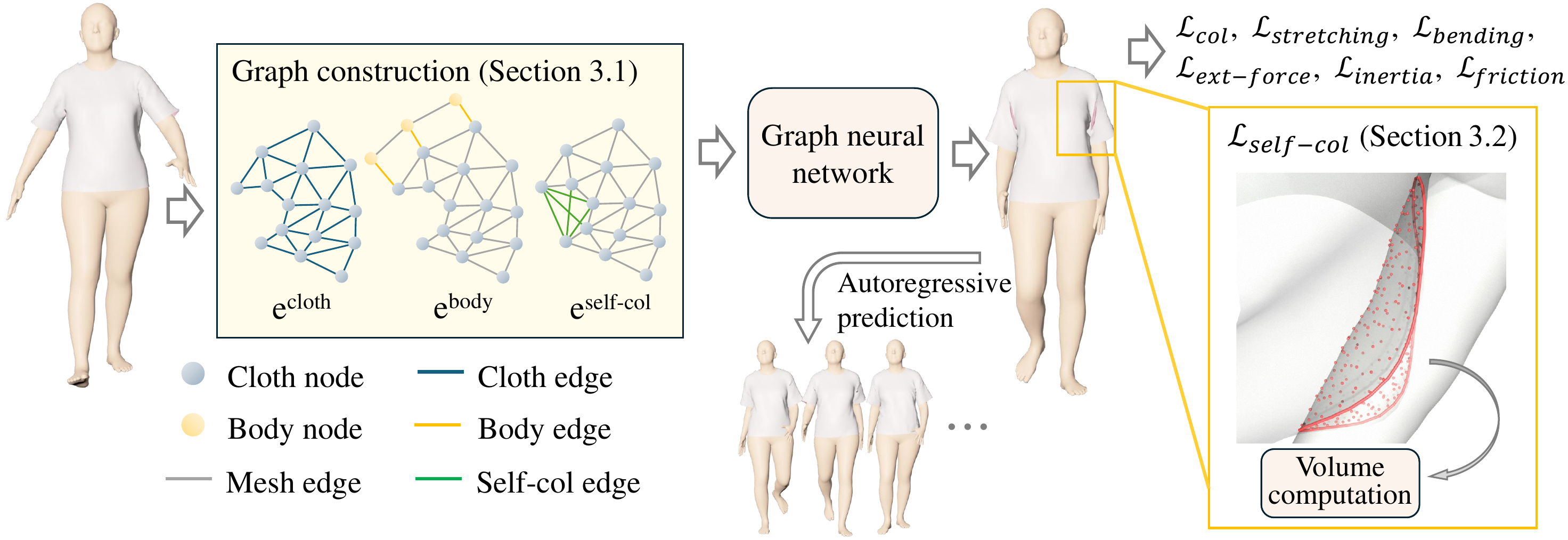}
  % \vspace{-0.2cm}
  \caption{Method overview.}
  \label{fig:overview}
  % \vspace{-0.4cm}
\end{figure}

\section{Methodology}
We aim to learn a neural model that autoregressively predicts the dynamics of the cloth, which should be physically correct and have minimal collision, including cloth self-collision and cloth-body collision.
We address the cloth self-collision issue in neural cloth simulation, by introducing the self-collision-aware network and the self-collision loss. The overview of our method is shown in \figref{overview}.
In this section, we first describe our self-supervised neural cloth simulator in \secref{simulator} and then about the 
the self-collision loss in 
\secref{compute_self}, which is the key to handling cloth self-collision in our framework.

%Among all energy terms, self-collision loss is the key to handling cloth self-collision. Thus, in Section~\ref{sec:compute_self}, we will introduce our self-collision loss in detail.

\subsection{Self-Collision-Aware Neural Cloth Simulator}
\label{sec:simulator}
In this section,  we first introduce our problem settings and the models we build upon. Next, we discuss why previous network structures cannot be used to handle the self-collision problem and present the self-collision-aware graph neural network. Finally, we describe our physical energy model and how it is used to supervise the training without ground truth data.

% \paragraph{Naming Convention}
% Lowercase bold letters denote a single vector, and uppercase bold letters denote a matrix. The regular font denotes a scalar.

\subsubsection{Background}
\label{sec:background}
Following MeshGraphNets~\cite{pfaff2020meshgraphnet} and HOOD~\cite{grigorev2023hood}, our model is a neural cloth simulator that predicts the state of the cloth mesh at time $t+1$ given the current state at $t$.
The cloth and the body are both represented by the mesh $M = \{ \VV, \FF\}$ with vertices $\VV$ connected as faces $\FF$. A graph neural network, which is agnostic to mesh connectivity, is used to process the cloth and its interaction with the body.

% how the graph is constructed. refer to Section 3.1 in HOOD, Basic Structure and Extension for clothing
The graph neural network used in MeshGraphNets~\cite{pfaff2020meshgraphnet} and HOOD~\cite{grigorev2023hood} contains two types of features: nodal feature $\xx$ and edge feature $\ee$.
The nodal feature represents the state of vertices.
The input nodal features to the network consist of the vertex type (body or cloth), velocity, normal, and physical material properties.
The edge feature characterizes the interaction between two vertices.
Different types of edges exist, including the cloth-cloth edge $\ec$ and cloth-body edge $\eb$.
The input cloth-cloth edge features consist of the relative position of vertices $\vv_i - \vv_j$ and the norm $|\vv_i - \vv_j|$ in both the current and canonical states.
The cloth-body edge connects every cloth vertex to the closest body vertex if their distance is below a threshold.
Its input features contain the relative position and the distance in the current and the previous frame.

% talk about the network itself. refer to Section 3.1 and 3.2 in HOOD
The input nodal and edge features are first embedded into latent vectors, followed by several hierarchical message-passing blocks~\cite{grigorev2023hood}.
In each message-passing, each type of edge features are first independently processed by different networks, and then node features are updated by incorporating its incident edge features:
\begin{equation}
    \ee'_{ij} \leftarrow f_{\text{edge}}(\ee_{ij}, \vv_i, \vv_j), 
    \quad \quad
    \xx'_{i} \leftarrow f_{\text{node}}(\xx, \sum_j \ee'^{\text{ body}}_{ij}, \sum_j \ee'^{\text{ cloth}}_{ij})
\end{equation}
All networks here are multi-layer perceptrons (MLP).
The last layer is an additional MLP that converts latent features into vertex accelerations, from which the vertex positions of the next frame can be obtained using the explicit Euler integration method.

\subsubsection{Self-collision-aware Graph Neural Network}
\label{sec:sca-gnn}
% \zhou{why original hood can't handle self-collision well}
Although the graph neural network in HOOD~\cite{grigorev2023hood} is able to produce clothing dynamics with minimal body-cloth collision, it cannot be used to handle cloth self-collision (see Section~\ref{sec:abl_study}).
Their networks use cloth mesh connectivity to reconstruct the graph, and the messages are propagated topologically, which is similar to how the local deformation of the cloth propagates across the mesh following its topological connections.
However, in many cases, the self-collision occurs between two parts of the cloth that are topologically far (see Figure~\ref{fig:torus_four} (c) and (d)).
\par
% \zhou{construction of cloth-cloth edge. an equation}
Based on this observation, we construct additional edges between cloth vertices according to their spatial distances.
For each cloth vertex $\vv_i$, we search for vertex $\vv_j$ so that $\| \vv_i - \vv_j \| < r$, and construct a self-collision edge $\esc_{ij}$ between them. 
$r$ is set to 2 cm empirically. 
Moreover, We exclude all original mesh edges when constructing self-collision edges because if two vertices share a mesh edge, they will not collide with each other.
In our ablation study, we found excluding the original mesh edges does not harm the model performance while saving computation.
\par
The self-collision edges are updated similarly to other types of edges.
The nodal features update becomes:
\begin{equation}
\label{eq:sel-col-gnn}
    \xx'_{i} \leftarrow f_{\text{node}}(\xx, \sum_j \ee'^{\text{ body}}_{ij}, \sum_j \ee'^{\text{ cloth}}_{ij}, \sum_j \ee'^{\text{ self-col}}_{ij})
    .
\end{equation}
\par
% \zhou{construction of external forces}
Additionally, we model the variable external forces by appending the external force to the input nodal feature.
During training in each iteration, we generate a force of random magnitude and direction, add it to the gravity, and set it as the input nodal feature.
After training, our model enables the user to provide external forces, such as the wind, for more dynamic behaviors.

\subsubsection{Cloth Energy Model}
\label{sec:energy}
The cloth energy model is defined here to reflect the physical properties of the cloth and can be used to supervise the model training without ground-truth data.
The self-collision term $\mathcal{L}_{self\text{-}col}$ penalizes the self-penetration volume, therefore preventing self-collision of the garment, as will be explained in Section~\ref{sec:compute_self}. 
The body-cloth collision term $\mathcal{L}_{col}$ is \textbf{max}$(\epsilon - \textbf{SDF}(x), 0)$, which measures the signed distance function for every vertex of the garment with respect to the body mesh, therefore, prevents the collision between the garment and the body, as explained in \cite{santesteban2021self}. 
The stretching term models the stretching forces with the St.Venant-Kirchhoff material \cite{montes2020computational}. 
Similarly, the bending term \cite{grinspun2003discrete} models the resistance to deformations that attempt to misalign adjacent faces. 
The external force energy for vertex $\vv_i$ is $-\mathbf{q}_i\vv_i$, where $\mathbf{q}_i$ is the sum of external forces except contact forces caused by the body.
The inertia term \cite{grigorev2023hood} represents inertia, which means objects tend to maintain their original velocities. 
The friction term models the friction forces between the garment and the body, as described in \cite{grigorev2023hood, brown2018accurate, geilinger2020add}.
The full loss is

\begin{equation}
\begin{aligned}
    \mathcal{L} & = \mathcal{L}_{self\text{-}col} (\VV^{t+\Delta t})
    + \mathcal{L}_{col} (\VV^{t}, \VV^{t+\Delta t})
    + \mathcal{L}_{stretching}(\VV^{t+\Delta t})
    + \mathcal{L}_{bending}(\VV^{t+\Delta t}) \\
    & + \mathcal{L}_{ext\text{-}force}(\VV^{t+\Delta t})
    + \mathcal{L}_{inertia} (\VV^{t-\Delta t}, \VV^{t}, \VV^{t+\Delta t})
    + \mathcal{L}_{friction} (\VV^{t+\Delta t})
    .
\end{aligned}
\end{equation}

\begin{algorithm}[t]
\caption{Compute the penetration volume}
\label{alg:compute_volume}
\begin{flushleft}
        \textbf{Input:} $\VV, \FF, \bm{\mathcal{B}}$ 
        \Comment{Vertices and faces of the garment, and the hole boundaries}
        \\
        \textbf{Output:} $volume$
\end{flushleft}
\begin{algorithmic}[1]
% \State $j \gets -1$
% \State Initialize as \hyperref[alg:reinit]{Reinitialize$()$}
\State $volume \gets 0$;
\State $\VV_\text{cls}, \FF_\text{cls} \gets \VV, \FF$;
\Comment{Initialize $\VV_\text{cls}, \FF_\text{cls}$}
\For{each boundary path $b$ in $\bm{\mathcal{B}}$}
\Comment{Close the garment}
\State $v_\text{central} \gets$ \textbf{MeanBoundary}($b$)
\State Add $v_\text{central}$ to $\VV_\text{cls}$;
\State Connect $v_\text{central}$ with vertices on $b$ and add faces to $\FF_\text{cls}$;
\EndFor{}
\State $\VV_\text{re}, \FF_\text{re} \gets$ \textbf{Remesh}($\VV_\text{cls}, \FF_\text{cls}$);
\Comment{Remesh the garment}
\State $\bm{\mathcal{P}} \gets$ \textbf{FindIntersectionPath}($\VV_\text{re}, \FF_\text{re}$);
\Comment{Global Intersection Analysis starts}
\For{each path $p$ in $\bm{\mathcal{P}}$}
\State $\bm{\mathcal{F}}_\text{pen}$ += \textbf{ParallelFloodFill}($p, \FF_\text{re}$);
\EndFor{}
\State $volume \gets $ \textbf{ComputeVolume}($\bm{\mathcal{F}_\text{pen}}, \VV_\text{re}$);
\Comment{Compute the penetration volume}

\end{algorithmic}
\end{algorithm}

\subsection{Self-collision Loss}
\label{sec:compute_self}
% \sinan{the part about self-collision is too long, consider separating it as a single section}
In this section, we explain the process to compute the self-collision loss that is based on the penetration volume that is formed when the garment intersects with itself (see  Figure~\ref{fig:torus_four}). The algorithm is outlined in Alg. \ref{alg:compute_volume}, where input $\VV$ and $\FF$ are the vertices and faces of the garment that we want to compute the self-intersection loss, and $\bm{\mathcal{B}}$ is the hole boundaries. %, as will be explained in \ref{sec:garment_closure}. 
The process of computing the self-collision loss can be divided into the following steps that we describe next: 
(1) {\bf garment closure} (steps 2-7), (2)  {\bf remeshing} (step 8), (3) {\bf Global Intersection Analysis} (steps 9-12), and (4) {\bf computing the penetration volume} (step 13).
%(1) garment closure (steps 2-7, \secref{garment_closure}), (2)  remeshing the garment (step 8, \secref{remesh}), (3) Global Intersection Analysis (steps 9-12, \secref{gia}), and (4) computing the penetration volume (step 13, \secref{penetration_volume}).

\subsubsection{Garment Closure}
\label{sec:garment_closure}
\begin{figure}[tb]
  \centering
  \includegraphics[width=0.52\textwidth]{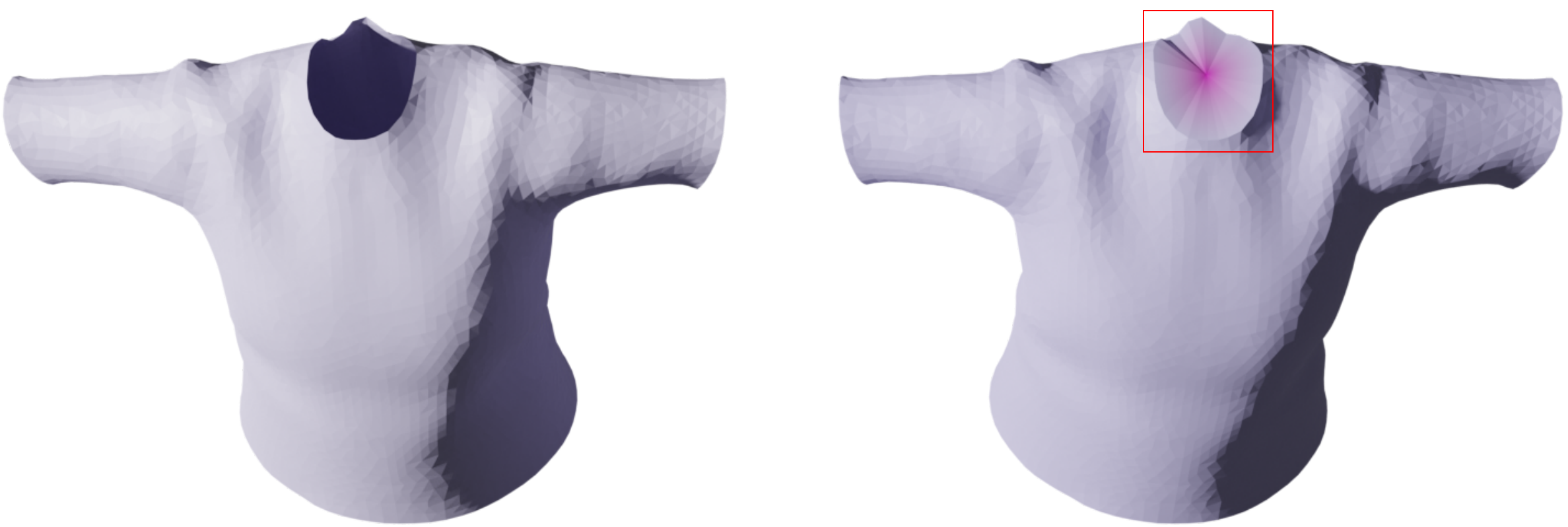}

  \caption{An example showing how to close the garment.
  }
  % \vspace{-0.5cm}
  \label{fig:close_garment}
\end{figure}

We first produce a closed surface of the garment by filling in the holes, such as the cuff of the shirt, so that the penetration parts form a closed mesh for computing the volume (see \figref{close_garment}). 
% Here we assume that the topology of the garment is fixed during the simulation. 
This is done by simply computing the average position of each hole boundary 
(represented by $v_\text{central}$ in Alg. \ref{alg:compute_volume})
and adding fan triangles that connect the mean point 
$v_\text{central}$
and each edge of the hole boundary.  
This process is conducted for all 
the hole boundaries
(represented by $\bm{\mathcal{B}}$ in Alg. \ref{alg:compute_volume}).
Regarding garments with more complex geometry, such as shirts with collars, we may choose to leave certain openings unclosed. Despite these remaining openings, our model is capable of learning to eliminate self-intersections from other parts, and still well handles these garments during inference, thanks to the generalization ability of our GNN. Alternatively, a more complex approach is required to define how to close the garment. This method must ensure that the addition of new triangles does not introduce non-existent self-collisions, thereby restricting the garment's movement. Addressing this challenge is designated as part of our future work.%We exclude garments that do not have manifold structures in this paper.   

 \subsubsection{Remeshing}
\label{sec:remesh}
\begin{figure}[tb]
  \centering
  \includegraphics[width=0.8\textwidth]{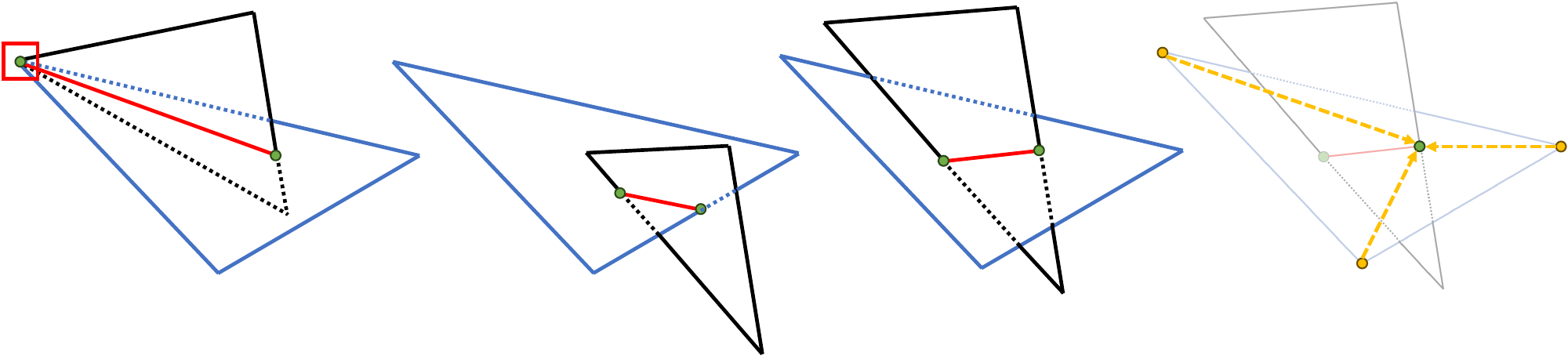}
  % \vspace{-0.2cm}
  \caption{The leftmost picture shows the case of the loop vertex \cite{baraff2003untangling} (enclosed by a red box), where one intersection point is the vertex shared by the two triangles. The middle two pictures show the other cases of two triangles intersecting, generating two intersection points (green circles) respectively. The two green points in these three cases become neighbors in the intersection path (the red line is a segment of the intersection path). These three cases are all the possible cases of two triangles intersecting. The right-most picture shows one intersection point can be represented by the three vertices (yellow circles) of the face using barycentric coordinates.
  }
  % \vspace{-0.5cm}
  \label{fig:triangle_inter}
\end{figure}

Next, given a configuration with self-intersection, we remesh the geometry of the garment so that all intersections lie exactly on the edges of the remeshed garment. 
%The garment is first remeshed as \sinan{cite libigl remesh}\cite{libigl} so that all intersections lie exactly on edges of the new mesh. 
The remeshed vertices and faces are represented as $\VV_\text{re}$ and $ \FF_\text{re}$ in Alg. \ref{alg:compute_volume}. All intersection points are uniquely recorded by $\left(edge, face\right)$ pairs, and the exact position of the intersection point can be expressed by the barycentric coordinates of the face. 
When two triangles intersect, the newly added two intersection points become neighbors in the intersection path, as shown in Figure~\ref{fig:triangle_inter}.
 The detection of triangle-triangle intersection and remeshing process are done using the functions of libiGL library~\cite{libigl}.

 \subsubsection{Global Intersection Analysis (GIA)}
\label{sec:gia}
\begin{figure}[tb]
  \centering
  \includegraphics[width=0.65\textwidth]{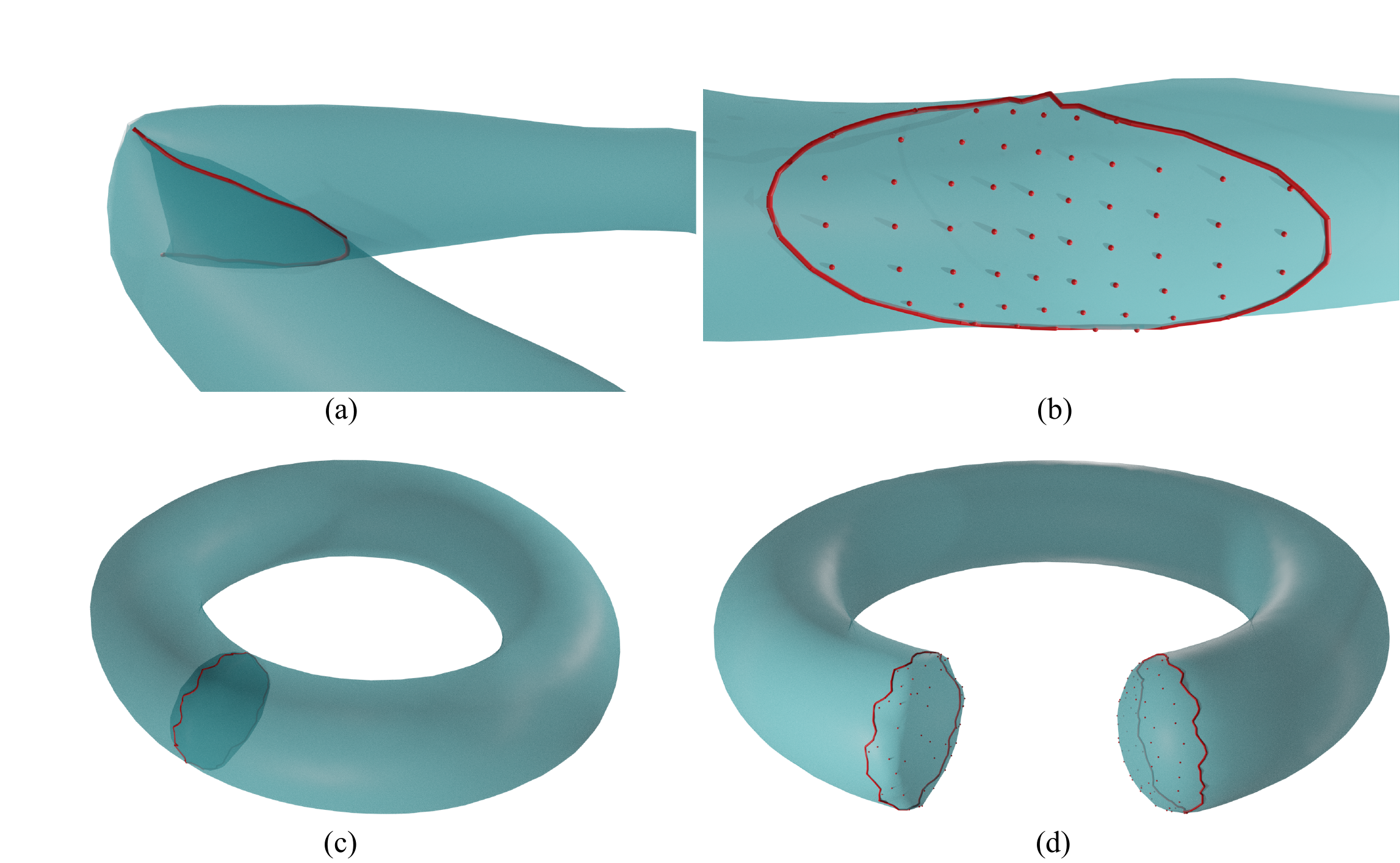}

  \caption{Here we show two cases of self-collision, 
  where the penetration volume is composed of one (a)(b) and two (c)(d) interaction paths.  
   (a) and (b) show a severely bent elbow with self-collisions happening inside the elbow. Figure (a) shows the intersection path. Figure (b) shows the vertices inside the self-collision and the intersection path after unbending the arm. Similarly, (c) and (d) show a case where a torus intersects with itself, resulting in two intersection paths and two separate penetration surfaces.
  }
  \label{fig:torus_four}
\end{figure}

In this section, we describe how we compute the paths of edges formed by the self-intersection (red lines in \figref{torus_four}, denoted as intersection paths here), and the set of faces surrounded by them.  
Following \cite{baraff2003untangling}, we find all intersection paths (represented by $\bm{\mathcal{P}}$ in Alg. \ref{alg:compute_volume}) by starting from any $\left(edge, face\right)$ intersection pair and tracking its neighbors using the information collected during remeshing. 
There are two types of self-intersections: those forming only one (\figref{torus_four}(a),(b)) intersection path and those forming two(\figref{torus_four}(c),(d)). The former case happens when a single region has been folded on top of itself while the latter is produced when two distinct regions of the mesh intersect. The key difference between these two cases is the existence of loop vertices. A loop vertex is defined as the vertex shared by two intersecting triangles, as shown in \ref{fig:triangle_inter}. The former can be distinguished from the latter when we find the loop vertices exist in the intersection paths. Loop vertices can be identified during the intersection tests of triangles because we can only find one $\left(edge, face\right)$ intersection pair when a loop vertex exists, as shown in the left most case of two triangles intersecting in Figure ~\ref{fig:triangle_inter}.
% The latter can be distinguished from the former when tracking the intersection path as it will end with a  $\left(edge, face\right)$ pair without looping back to the one started from.

Next, the set of faces bound by the intersection path 
that forms the penetration volume (represented by $\bm{\mathcal{F}}_\text{pen}$ in Alg. \ref{alg:compute_volume}) are extracted. 
Since it's topologically ambiguous which set of faces bound by the intersection path is in penetration, 
e.g., for Figure~\ref{fig:torus_four} (b), %the system doesn't know which side is in penetration, 
the one enclosed by the intersection path or the rest of the body. Therefore, a heuristic is formed that the smaller side is the one in penetration. 
%By "smaller", we mean the side which has less triangles. Here we assume the areas of the faces are more less the same. 
The parallel flood fill algorithm is then used, where we traverse the faces on both sides of the intersection path simultaneously while prohibiting the traversal through an edge on the intersection path. The side that first finishes the traversal is considered to be in the penetration region. For those intersection paths formed by two distinctive regions (the bottom two in Figure~\ref{fig:close_garment}), we need to traverse twice to obtain two groups of faces in penetration. 
For those intersection paths formed by one region
 (the top two in Figure~\ref{fig:close_garment}), the penetration faces are enclosed by one intersection path, so one traversal can extract the mesh forming the intersection. We refer readers to \cite{baraff2003untangling}
 for more details of GIA.

 \subsubsection{Computing the Penetration Volume}
\label{sec:penetration_volume}
Using the faces that form the penetration volume extracted by GIA, we compute the penetration volume that is used to compute the self-collision loss in \secref{energy}. 
The volume of the closed mesh can be computed by summing the signed volume, i.e., the scalar triple product of every tetrahedron that is composed of a face in penetration and the origin. Since all the new vertices produced at the remeshing step are represented using barycentric coordinates of the original vertices, the volume loss can then be backpropagated through the neural network.

\section{Experiments}
\label{sec:exp}
\begin{table}[t]
\centering
\renewcommand{\arraystretch}{1.1}
\setlength{\tabcolsep}{2pt} 
\begin{tabular}{|c|ccc|ccc|}
\hline
     & \multicolumn{3}{c|}{t-shirt}                                           & \multicolumn{3}{c|}{skirt}                                             \\ \cline{2-7}
 &
  \multicolumn{1}{c|}{\makecell{$\mathcal{L}_{self\text{-}col}$ \\ ($\times 10^{-3} $) $\downarrow$}} &
  \multicolumn{1}{c|}{\% (0.1) $\downarrow$} &
  \% (0.01) $\downarrow$ &
  \multicolumn{1}{c|}{\makecell{$\mathcal{L}_{self\text{-}col}$ \\ ($\times 10^{-3} $) $\downarrow$}} &
  \multicolumn{1}{c|}{\% (0.1) $\downarrow$} &
  \% (0.01) $\downarrow$ \\ \hline
% SNUG
SNUG  & \multicolumn{1}{c|}{52.00}        & \multicolumn{1}{c|}{24.78}        &  38.76        &  \multicolumn{1}{c|}{N/A} &  \multicolumn{1}{c|}{N/A}& N/A\\ 
% NCS
NCS  & \multicolumn{1}{c|}{75.05}        & \multicolumn{1}{c|}{33.24}        &  70.58        & \multicolumn{1}{c|}{476.03}         & \multicolumn{1}{c|}{99.91}        & 100        \\ 
HOOD & \multicolumn{1}{c|}{26.10} & \multicolumn{1}{c|}{8.92} & 22.62 & \multicolumn{1}{c|}{7.33} & \multicolumn{1}{c|}{1.52} & 9.84 \\ \hline
% ablation 1
\makecell{w. area loss} &
  \multicolumn{1}{c|}{28.81} &
  \multicolumn{1}{c|}{8.18} &
   39.08 & 
  \multicolumn{1}{c|}{8.80} &
  \multicolumn{1}{c|}{1.93} &
  13.61
   \\ 
% ablation 2
\makecell{ w/o self-col edge} &
  \multicolumn{1}{c|}{\textbf{1.14}} &
  \multicolumn{1}{c|}{\textbf{0}} &
  \textbf{1.70} &
  \multicolumn{1}{c|}{7.92} &
  \multicolumn{1}{c|}{1.61} &
  12.69 \\  
% ablation 3
\makecell{ w. mesh edge} &
  \multicolumn{1}{c|}{2.37} &
  \multicolumn{1}{c|}{\textbf{0}} &
  5.47 &
  \multicolumn{1}{c|}{2.02} &
  \multicolumn{1}{c|}{\textbf{0.14}} &
  3.95 \\ 
   \hline
% final
SENC & \multicolumn{1}{c|}{1.80} & \multicolumn{1}{c|}{\textbf{0}}       & 3.724  & \multicolumn{1}{c|}{\textbf{1.59}} & \multicolumn{1}{c|}{0.28} & \textbf{2.71} \\ \hline
\end{tabular}
\caption{Quantitative comparison with SOTA methods (upper part) and ablation study (lower part).}
% \vspace{-0.7cm}
\label{tab:comparison}
\end{table}

\subsubsection{Implementation Details}
Following \cite{santesteban2022snug,grigorev2023hood}, we use 52 human motion capture sequences from the AMASS~\cite{mahmood2019amass} dataset for training.
Our training consists of two phases: pre-training without self-collision loss and full training with all losses.
At the beginning of training, the network output is very noisy, and conducting GIA for such data is very time-consuming. 
Numerical errors could also happen during the process of remeshing and GIA, especially when numerous triangles gather together.
Thus, we pre-train the network without the self-collision loss for 120,000 iterations and then train it for 70,000 iterations with all losses.
The whole training process takes around 48 hours on a single Nvidia RTX 4090.
We borrow several training techniques from HOOD~\cite{grigorev2023hood}, including initializing garment and autoregressive training.

\subsubsection{Competing Methods}
We compare SENC with several recent representative neural cloth simulation methods.
\textbf{SNUG}~\cite{santesteban2022snug} is one of the pioneering works that learn cloth dynamics in a self-supervised manner.
It prevents body-cloth collision by introducing a collision loss based on the human body SDF.
However, since the authors do not release the training code, we can only use their pre-trained models for certain types of garments.
Similarly, \textbf{NCS}~\cite{bertiche2022neural} is trained in a self-supervised manner, while they claim to have better dynamics than previous works.
\textbf{HOOD}~\cite{grigorev2023hood} is the first self-supervised neural cloth simulator to model clothing dynamics for arbitrary mesh topology and connectivity.
However, none of them handles cloth self-collision, which leads to a great number of artifacts.

\begin{figure}[t]
  \centering
  \includegraphics[width=0.9\textwidth]{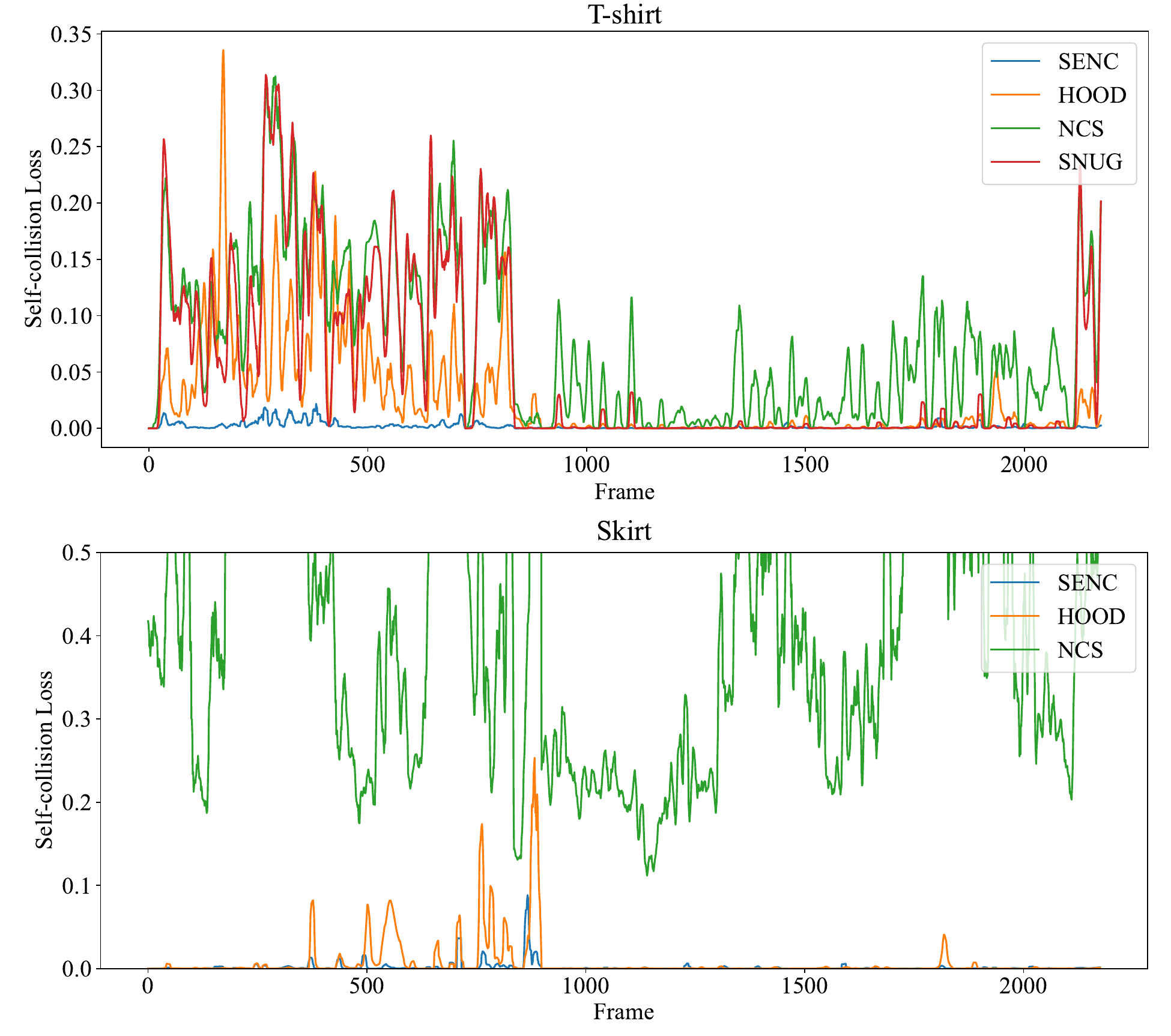}
  \caption{Quantitative comparisons showing the self-collision loss of different methods on the test sequences. Our method (SENC) shows a significantly lower loss.
  }
  % \vspace{-0.5cm}
  \label{fig:curve}
  % \vspace{-0.1cm}
\end{figure}

%%%%%%%%%%%%%%%%%%%%%%%%%%%%%%%%%%%%%%%%%%%%%%%%%%%%%%%%%%%%%%
\subsection{Quantitative Evaluation}
As in \cite{santesteban2022snug}, We evaluate our model on 4 test sequences from AMASS~\cite{mahmood2019amass} containing 2175 frames, which are unseen during training.
Since SNUG and NCS do not support physical material control, we set the same set of fabric materials for all experiments.
We analyze the cloth self-collision using the average self-collision loss $\mathcal{L}_{self\text{-}col}$ computed over all test frames.
In addition, we calculate the percentage of frames whose $\mathcal{L}_{self\text{-}col}$ is above a certain threshold. 
We use $\% (0.1)$ and $\% (0.01)$ to denote the percentages with threshold of $0.1$ and $0.01$ respectively.
\par
In Table~\ref{tab:comparison}, we first compare our method with SNUG, NCS, and HOOD.
It can be seen clearly that our method outperforms all other methods by a great margin.
As the authors of SNUG did not release the training code, and they do not have the checkpoint for the skirt, we skip its evaluation on the skirt.
\par
Figure~\ref{fig:curve} visualizes the self-collision losses on the test sequences.
This demonstrates that our method successfully addresses the issue of cloth self-collision, in contrast to other approaches that frequently result in such problems.

\subsubsection{Ablation Study}
\label{sec:abl_study}
We further conduct an ablation study to validate the effectiveness of our designs.
\textbf{w. area loss} denotes that we compute the area of the intersection part (summing all the areas of the faces inside the penetration) instead of the volume.
\textbf{w/o self-col edge} is the model without constructing the self-collision edges in Equation~\ref{eq:sel-col-gnn}.
\textbf{w. mesh edge} means we keep all edges, including mesh edges when constructing the self-collision edges. 
In contrast, our final model excludes original mesh edges when constructing the self-collision edges, because if two vertices are neighbors (connected by a mesh edge), they will not collide with each other.
\par
From Table~\ref{tab:comparison}, we can see that if we use the area (\textbf{w. area loss}) instead of the volume, the model can hardly learn to handle self-collision.
%Thus, we choose to compute the volume for our self-collision loss.
%This could be because only the vertices near the intersection path affects the area and the gradients are zero for all the vertices inside the intersection path. 
This could be because the gradient to reduce the area of the penetration surface does not effectively direct the vertices to resolve the penetration.
\textbf{w/o self-col edge} well addresses the self-collision for the t-shirt.
However, it fails to handle the skirt.
Compared to the t-shirt, the collision part of the skirt has larger topological distances in the mesh graph.
For example, the front and rear hem, which are far topologically from the skirt, could easily collide.
With the help of self-collision edges, such self-collision can be well addressed, leading to a lower self-collision loss of our final method.
\textbf{w. mesh edge} has a slightly worse performance than our final method, showing that including the original mesh edges in self-collision edges does not boost the performance but only introduces extra computation.

\begin{figure}[tb]
  \centering
  \includegraphics[width=0.9\textwidth]{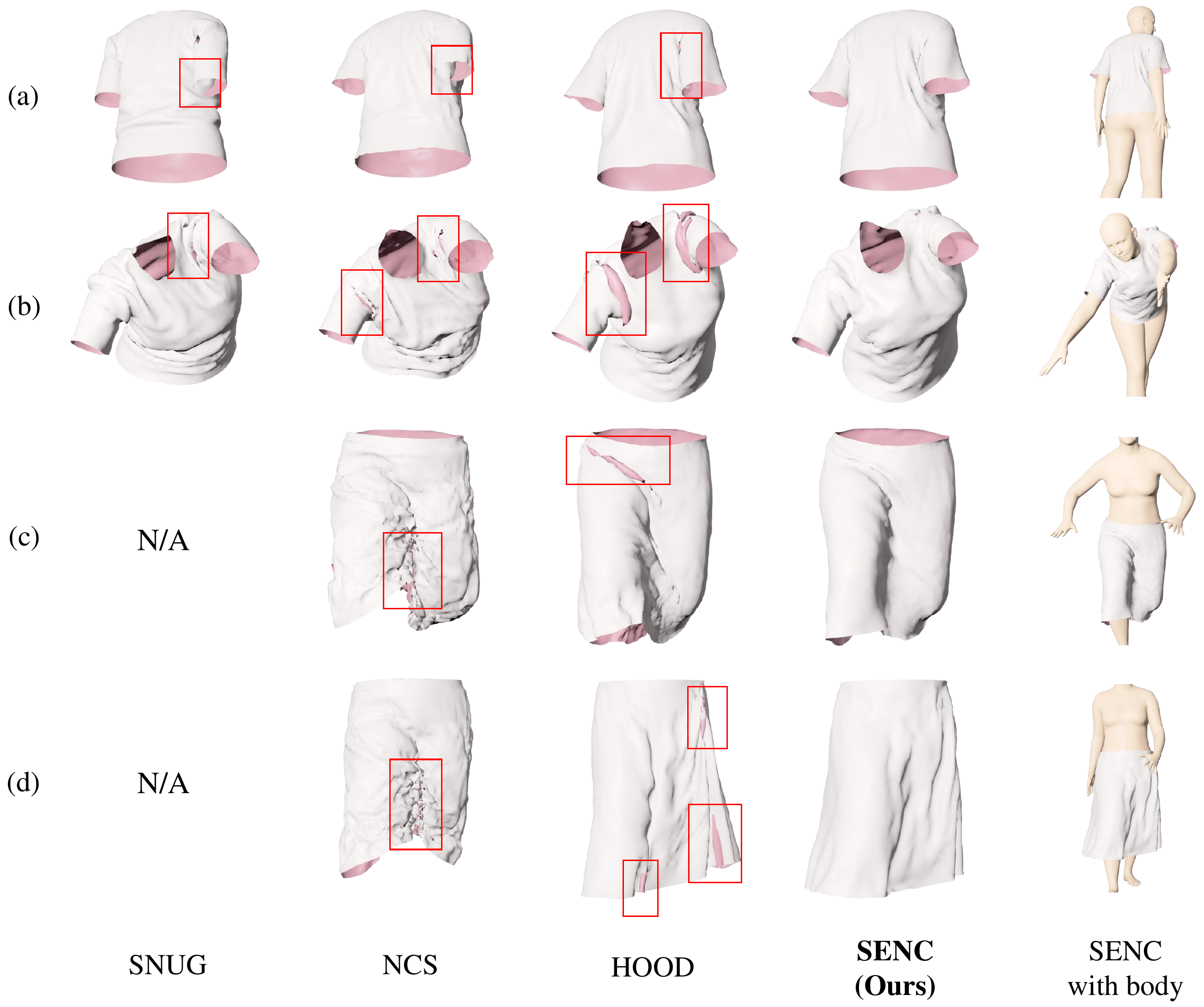}
  % \vspace{-0.3cm}
  \caption{Comparisons with existing methods. Other methods exhibit clear cloth self-collision, while our method addresses it well.}
  % \vspace{-0.3cm}
  \label{fig:comparison}
\end{figure}

%%%%%%%%%%%%%%%%%%%%%%%%%%%%%%%%%%%%%%%%%%%%%%%%%%%%%%%%%%%%%%
\subsection{Qualitative Evaluation}
We visualize the results of our method and other competing methods in Figure~\ref{fig:teaser} and Figure~\ref{fig:comparison}.
To enhance the clarity of our visual results, we render the outer side of the garments in white and the inner side in pink.
In Figure~\ref{fig:comparison} (a), SNUG and NCS both have a severe self-collision, with a notable portion of the t-shirt sleeve intruding into the torso region of the garment.
In (b), all competing methods have a clear self-collision around the shoulder.
HOOD has a much larger penetration volume than SNUG and NCS as it is a GNN-based model and has better dynamics.
Our method, despite also being GNN-based, avoids the penetration successfully.
In (c) and (d), NCS produces a great number of artifacts, as it is a skinning-based pose-dependent method.
Similarly, HOOD is still not immune to noticeable self-collision problems.
\par
\begin{figure}[tb]
  \centering
  \includegraphics[width=0.6\textwidth]{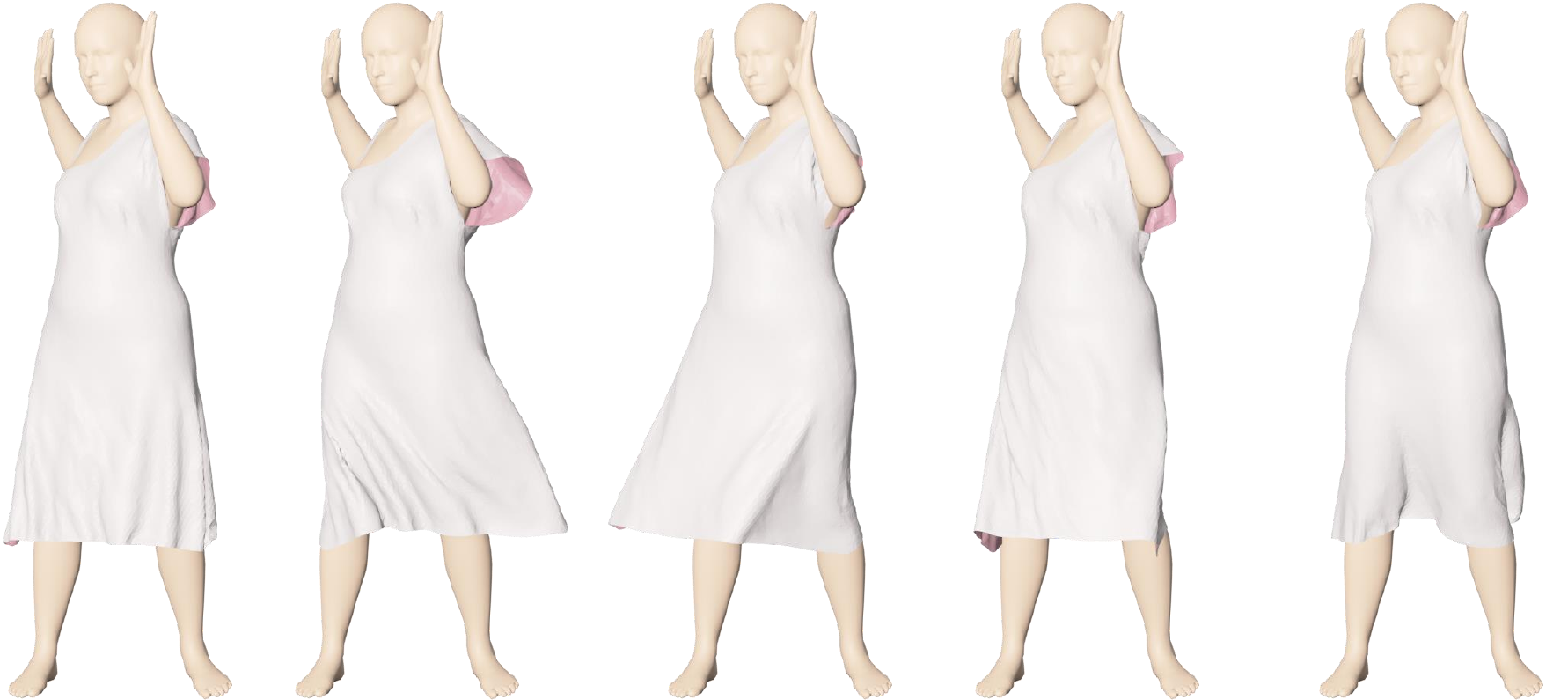}
  \caption{Our method enables variable external forces on the cloth. Here the dress is depicted responding to wind from various directions, with the leftmost one not affected by external forces except gravity.
  }
  \label{fig:wind}
  % \vspace{-0.3cm}
\end{figure}

In Figure~\ref{fig:wind}, we demonstrate that our model enables the user to apply a variable external force on the cloth.
The cloth deforms naturally since we jointly optimize the energy by the external force and other energies.

\section{Conclusion}
We propose a self-supervised scheme for Neural Cloth Simulation, which solves the persisting problem in the literature: self-collision. 
%Due to two reasons, previous methods are not able to solve this problem. The first is that existing collision handling techniques cannot be easily integrated into the framework, because they either can only prevent collisions before they occur, or they demand post-process and cannot be directly integrated. The second is that garments are cloth-like thin structures, which means they do not have an explicit definition for the penetration side when self-collisions happen. 
%Therefore, 
Based on GIA proposed in \cite{baraff2003untangling}, we have developed a new self-collision loss, which 
%effectively solves this problem and 
can be easily integrated into any existing neural cloth simulation framework, without sacrificing the high quality of garment dynamic shown in the state-of-the-art\cite{grigorev2023hood}. 

\subsubsection{Limitations and Future Work}
While our method is able to address self-collisions effectively, it still has some limitations. Firstly, computing the penetration volume is time-consuming, due to bottlenecks caused by remeshing and GIA.
% Accelerating such processes through parallelization on the GPU is a possible future research direction.
Better algorithms for accelerating these processes are a possible future research direction.
Secondly, we are only able to handle meshes that can be easily closed. For garments with complex topology, it may be difficult to define how to close the garment. The dilemma is described as follows. Take the long skirt as an example: closing the garment can sometimes inhibit valid deformation. If the bottom hole is closed, then our model will prevent the bottom part of the skirt from moving upwards, as it can result in producing penetration volume between the virtually added triangles and the original skirt mesh. Conversely, leaving it open means the model cannot learn from the cases when self-collisions happen between the front and back hems, as these penetrations will not be closed without sealing the bottom hole.
A penetration loss that can roughly compute the bounded volume without closing the boundary is desired: we can possibly apply winding numbers~\cite{Barill:FW:2018} or electronic flux~\cite{wang2013harmonic} for this purpose.   
%Therefore, handling self-collisions on the boundaries of the garment is still a remaining problem for future work.
Moreover, our method can generalize to other objects such as multi-layer cloth and 3D deformable characters: extending our method to such topics would be promising.

\section*{Acknowledgement}
This work is partly supported by the
Innovation and Technology Commission of the HKSAR Government under  the ITSP-Platform grant 
(Ref: ITS/335/23FP) and
the InnoHK initiative (TransGP project), 
The research work was in part conducted in the JC STEM Lab of Robotics for Soft Materials funded by The Hong Kong Jockey Club Charities Trust.
\newpage
\appendix
\begin{center}
\textbf{\large SENC: Handling Self-collision in Neural Cloth Simulation}
\\
\textbf{\large --Appendix-- }
\end{center}

\section{Repulsive Loss}
\begin{table}[]
\centering
\renewcommand{\arraystretch}{1.1}
\setlength{\tabcolsep}{2pt} 
\begin{tabular}{|c|ccc|ccc|}
\hline
     & \multicolumn{3}{c|}{t-shirt}                                           & \multicolumn{3}{c|}{skirt}                                             \\ \cline{2-7}
 &
  \multicolumn{1}{c|}{\makecell{$\mathcal{L}_{self\text{-}col}$ \\ ($\times 10^{-3} $) $\downarrow$}} &
  \multicolumn{1}{c|}{\% (0.1) $\downarrow$} &
  \% (0.01) $\downarrow$ &
  \multicolumn{1}{c|}{\makecell{$\mathcal{L}_{self\text{-}col}$ \\ ($\times 10^{-3} $) $\downarrow$}} &
  \multicolumn{1}{c|}{\% (0.1) $\downarrow$} &
  \% (0.01) $\downarrow$ \\ \hline
% repulsive 
\makecell{repulsive loss} &
  \multicolumn{1}{c|}{15.16} &
  \multicolumn{1}{c|}{3.31} &
  29.93 &
  \multicolumn{1}{c|}{22.38} &
  \multicolumn{1}{c|}{2.30} &
  6.21 \\ 
   \hline
% ours
SENC & \multicolumn{1}{c|}{1.80} & \multicolumn{1}{c|}{0}       & 3.724  & \multicolumn{1}{c|}{1.59} & \multicolumn{1}{c|}{0.28} & 2.71 \\ \hline
\end{tabular}
\label{tab:comparison2}
\caption{Comparison with the repulsive loss.}
\end{table}

We further explore a possible solution, the repulsive loss~\cite{lee2023clothcombo}, to handle cloth self-collision.
It tries to separate non-adjacent cloth vertices when they are close, and can be formulated as:
\begin{equation}
    \mathcal{L}_{repulsive} = \sum_i^N \sum_{j \in \mathcal{A}_i} - \log (\vv_i - \vv_j)^2
    ,
\end{equation}
where $\mathcal{A}_i = \{ j \in V \mid (\vv_i, \vv_j) \notin E \text{ and } d(\vv_i, \vv_j) < \text{threshold} \}
$, and $d()$ is the distance function.
We set threshold = 5cm.

% describe the result in the table
Evaluated on the same sequences in Section 5 of our paper, the results in the table above additionally shows the comparisons between ours and the repulsive loss applied on HOOD\cite{grigorev2023hood}'s model. For each type of the garment (t-shirt or skirt), the left most column shows directly the averaged self-collision during the evaluation; the middle column shows the proportions of the frames whose self-collisions are higher than 0.1; Similarly, this threshold is set to 0.01 in the right column. It can be seen that ours outperforms the repulsive loss by a large margin.

One of the main reasons for its poor performance is that the repulsive loss simply prevents all vertices pairs from getting too close. Consequently, if self-collisions already exist, then it will also prevent penetrations from being resolved because it does not allow vertices in penetrations to approach the penetration surfaces from where they originally penetrated in.

\section{More Quantitative Results}
In Table~\ref{tab:more_garments}, we additionally show results on six more unseen garments from the test set of HOOD, where we present the self-collision loss of HOOD and our model on the test sequences.
Our model resolves self-collision significantly for all types of garments, further verifying the generalization ability and efficacy of our method.

\begin{table}[t]
\centering
\renewcommand{\arraystretch}{1.1}
\setlength{\tabcolsep}{8pt} 
\begin{tabular}{|c|c|c|c|c|c|c|}
\hline
 & pants  & short & tshirt & novel & hooded & tight \\
 & shorter & sleeve & dynamic & tank & dress & dress \\
\hline
HOOD & 6.08 & 60.66 & 164.23 & 6.73 & 33.37 & 35.86 \\
\hline
SENC & \textbf{0.25} & \textbf{2.90} & \textbf{6.58} & \textbf{0.13} & \textbf{2.63} & \textbf{1.56} \\
\hline
\end{tabular}
\caption{Comparison of $\mathcal{L}_\text{col}$ on more unseen garments}
\label{tab:more_garments}
\end{table}

\section{Runtime Speed}

We measure the runtime speed of our method and compare it with HOOD~\cite{grigorev2023hood} and \textbf{w. mesh edge}, an ablation setting where the mesh edge is not excluded when constructing self-collision edges (More details in the main paper).
The speed of our method is slightly slower than HOOD due to the construction of self-collision edges. However, our self-collision is significantly reduced compared to HOOD.
Compared to \textbf{w. mesh edge}, our final method has a faster speed and better self-collision prevention (See Table 1 of the main paper), which validates the effectiveness of our design.

\begin{table}[t]
\centering
\renewcommand{\arraystretch}{1.1}
\setlength{\tabcolsep}{8pt} 
\begin{tabular}{|c|c|c|c|c|}
\hline
     & t-shirt & skirt  & dress  & average \\ \hline
HOOD & 23.127  & 20.012 & 15.638 & 19.090  \\ \hline
w. mesh edge & 20.041  & 16.986 & 13.078 & 16.196  \\ \hline
SENC & 20.179  & 17.942 & 13.732 & 16.843  \\ \hline
\end{tabular}
\caption{Runtime speed (unit: frame per second).}
\label{tab:runtime}
\end{table}

\newpage

%\clearpage  % TODO REVIEW/FINAL: This \clearpage needs to be removed from both review and camera-ready versions.

\bibliographystyle{splncs04}
\bibliography{main}
\end{document}